\documentclass[conference]{IEEEtran}
\IEEEoverridecommandlockouts
\usepackage{cite}
\usepackage{amsmath,amssymb,amsfonts}
\usepackage{algorithmic}
\usepackage{algorithm}
\usepackage{graphicx}
\usepackage{textcomp}
\usepackage{xcolor}
\def\BibTeX{{\rm B\kern-.05em{\sc i\kern-.025em b}\kern-.08em
    T\kern-.1667em\lower.7ex\hbox{E}\kern-.125emX}}
\begin{document}

\title{Data Fusion and Aggregation Methods to Develop Composite Indexes for a Sustainable Future\\
{\footnotesize \textsuperscript{*}}
\thanks{Identify applicable funding agency here. If none, delete this.}
}

\author{\IEEEauthorblockN{Abdullah Konak}
\IEEEauthorblockA{\textit{Division of Engineering, Business, and Computing} \\
\textit{Penn State Berks}\\
Reading, PA, USA \\
https://orcid.org/0000-0001-6250-7825}
}

\maketitle

\begin{abstract}
Research on environmental risk modeling relies on numerous indicators to quantify the magnitude and frequency of extreme climate events, their ecological, economic, and social impacts, and the coping mechanisms that can reduce or mitigate their adverse effects. Index-based approaches significantly simplify the process of quantifying, comparing, and monitoring risks associated with other natural hazards, as a large set of indicators can be condensed into a few key performance indicators. Data fusion techniques are often used in conjunction with expert opinions to develop key performance indicators. This paper discusses alternative methods to combine data from multiple indicators, with an emphasis on their use-case scenarios, underlying assumptions, data requirements, advantages, and limitations. The paper demonstrates the application of these data fusion methods through examples from current risk and resilience models and simplified datasets. Simulations are conducted to identify their strengths and weaknesses under various scenarios. Finally, a real-life example illustrates how these data fusion techniques can be applied to inform policy recommendations in the context of drought resilience and sustainability.
\end{abstract}


\section{Introduction}

The growing challenges posed by climate change have highlighted the limitations of earlier drought risk assessment models, which focused narrowly on hazard characteristics such as the frequency and spatial extent of precipitation deficits, heatwaves, and other related factors. One critical limitation is that understanding the hazard dimension alone provides limited guidance for developing effective long-term adaptation strategies and policies. Consequently, more recent drought risk models have adopted a comprehensive approach that integrates the vulnerability of social, economic, physical, political, and environmental systems alongside the assessment of drought hazards \cite{UNDRR_2021}.

The Intergovernmental Panel on Climate Change (IPCC) defines vulnerability as the conditions and processes that increase the susceptibility of human and environmental systems to the impacts of natural disasters \cite{IPCC_2023}. Vulnerability has been conceptualized and integrated into risk assessment using a ``contextual-factor'' approach \cite{Birkmann_2013}, in which vulnerability is defined as a function of intrinsic system characteristics, represented by exogenous variables such as ``Exposure,'' ``Sensitivity,'' and ``Adaptive/Coping Capacity,'' as follows:

\begin{equation}
\text{Vulnerability} = f(\text{Exposure}, \text{Sensitivity}, \text{Adaptive Capacity})
\label{eq:vulnerability}
\end{equation}

In this formulation, exposure refers to the elements that are subject to the negative impacts of a hazard (e.g., population, infrastructure, and ecosystems) while sensitivity captures how severely those elements are affected. The potential impact, derived from the combination of exposure and sensitivity \cite{Fontaine_2009}, can be mitigated by adaptive capacity. Since the contextual-factor approach treats vulnerability as independent from hazard characteristics \cite{Brooks_2003}, the resulting metric reflects the potential impact on the system. Combining vulnerability with hazard, a relative measure of risk is defined as follows:

\begin{equation}
\text{Risk} = f(\text{Hazard}, \text{Vulnerability})
\label{eq:risk}
\end{equation}

Each component of the above risk function is evaluated using multiple indicators and indices. An \textit{indicator} refers to a direct measurement of a variable (e.g., GDP, rainfall). An \textit{index} aggregates multiple indicators into a single score to simplify complex information across time or space \cite{OECD2008}. Therefore, index-based approaches have become widely used in assessing and monitoring natural disaster risks and vulnerabilities.

Two essential steps in constructing a risk or vulnerability index are: (i) selecting relevant indicators and (ii) determining an appropriate aggregation and weighting method. Indicator selection is typically based on the context, objectives of risk assessment, and data availability \cite{Gonzalez2016, Sass2024}. The aggregation method determines how the selected indicators are combined, while the weighting method specifies the contribution of each indicator to the index. Despite the availability of alternative aggregation techniques, the weighted average approach is predominant in the literature.

Let $n$ independent indicators $X_1, \ldots, X_n$ be used to calculate an index $I_k$ to compare $m$ systems, where $k = 1, \ldots, m$, and $x_{ik}$ is the value of $X_i$ for system $k$. Assuming that higher values indicate desirable attributes (e.g., capacity or resilience), the index $I_k$ is defined as:

\begin{equation}
I_k = \sum_{i=1}^{n} w_i x_{ik}
\quad \text{subject to} \quad \sum_{i=1}^{n} w_i = 1
\label{eq:wa}
\end{equation}

Index developers must decide $w_i$ for each indicator in \eqref{eq:wa}. Weighting methods can be broadly categorized into groups: subjective and objective approaches. Subjective methods rely on expert judgment to assign weights based on the perceived importance of indicators. In contrast, objective methods determine weights from data, independent of expert input, using mathematical or statistical properties such as variance or entropy. Although subjective weights lack normative interpretation \cite{Schlossarek2019}, they are well-suited for comparing systems across contexts and provide replicable and transparent benchmarks.

While several objective methods have been applied in drought risk and resilience studies in the literature, many applications fail to justify their methodological choices. Consequently, it remains unclear how different weighting schemes affect index construction, particularly when indicators exhibit varying distributions or correlations. This paper addresses this gap by systematically evaluating a set of widely used objective weighting methods, including variance-based, entropy-based, PCA-based, CRITIC, and, for the first time in this context, Data Envelopment Analysis (DEA), to compare their behavior using simulation. 

The contribution of the paper is twofold: First, the sensitivity and performance of objective weighting methods are analyzed under controlled conditions with different statistical properties (e.g., correlated inputs, skewed distributions, systemic trends). Second, DEA is introduced as a novel method for indicator weighting in environmental risk modeling. The findings offer valuable insights for researchers and practitioners when selecting among objective methods for index construction in drought and other disaster risk applications.
 
\section{Objective Methods}
\subsection{Variance-based Weighting Method}
Assume that $n$ independent indicators $X_1, \ldots, X_n$ to measure a factor $I_f$. Since each indicator $X_i$ represents a different way of measuring factor $I_f$, the indicators may have different variances $\sigma_1^2, \ldots, \sigma_n^2$. This observation implies that the best estimator of $I_f$ can be obtained by the weighted mean of the indicators, as opposed to the arithmetic mean, as follows:

\begin{equation}
\hat{I}_f = \frac{\sum_{i=1}^{n} w_i x_i}{\sum_{i=1}^{n} w_i}
\label{eq:amean}
\end{equation}

where $w_i$ is the weight of $X_i$. Equation \eqref{eq:amean} has the smallest variance among all weighted averages when $w_i = \frac{1}{\sigma_i^2}$ for each $X_i$, which is called the inverse-variance weighting method in statistics. Note that the magnitude of $w_i$ does not imply the importance of $X_i$. Rather, the inverse-variance weighting method assigns proportionally lower weights to indicators with high variability. As a result, each additional indicator will only reduce $\text{Var}(\hat{I}_f)$ regardless of how noisy the additional indicator is, which in turn justifies using multiple indicators to measure a factor.

If $\sum\limits_{i=1}^{n}{{{w }_{i}}}=1$ is required, the optimal weights will be calculated as:

\begin{equation}
w_i = \frac{1}{\sigma_i^2 \sum_{j=1}^{n} \frac{1}{\sigma_j^2}}
\label{eq:varbased}
\end{equation}

Equation \eqref{eq:varbased}, therefore, provides a mechanism for combining multiple indicators into a single estimate, assigning each a weight proportional to its statistical reliability.

\subsection{Entropy-based Weighting Method}
Entropy is a measure of the level of ``information'' represented by a random variable's possible outcomes. The higher the entropy is, the more difficult it is to predict the outcomes of the random variable. For example, consider flipping a fair coin with $p(\text{`head'}) = 0.5$ and an unfair coin with $p(\text{`head'}) = 0.90$. We can predict the outcomes of flipping the unfair coin with more certainty than the outcomes of the fair coin. This means each observation from the unfair coin yields less information, on average, than one from the fair coin. Similarly, if we can calculate the entropies of indicators, we can claim that an indicator with higher entropy provides more information than another indicator with lower entropy. We can assume that an indicator $X_i$ can take $m$ discrete values $x_{i1}, x_{i2}, \ldots, x_{im}$ with corresponding probabilities $p(x_{i1}), p(x_{i2}), \ldots, p(x_{im})$.

The entropy of indicator $X_i$ is given by:
\[
H(X_i) = -\sum_{k=1}^{m} p(x_{ik}) \log( p(x_{ik}))
\]

Algorithm \ref{al:ent} \cite{Zhu2018} illustrates the calculation of weights using the entropy-based method. 

\begin{algorithm}
\caption{Entropy-Based Weighting Method}
\begin{algorithmic}[1]
\FOR{$i = 1$ to $n$} 
    \STATE Calculate $p(x_{ik})=\frac{x_{ik}}{\sum_{j=1}^{m}{x_{ij}}}$ for $k=1, \ldots,m$
    \STATE Set small constant $\varepsilon \gets 10^{-12}$ to avoid $\log(0)$
    \STATE $H(X_i) \gets - \frac{1}{\log{m}} \sum_{k=1}^{m} p(x_{ik}) \cdot \log(p(x_{ik}) + \varepsilon)$
\ENDFOR
\STATE $w_i \gets \frac{1-H(X_i)}{\sum_{i=1}^{n} (1-H(X_i))}$ 
\RETURN $\{w_1, w_2, \ldots, w_n\}$
\end{algorithmic}
\label{al:ent}
\end{algorithm}

\subsection{Principal Component Analysis}
In real-world applications, some drought indicators are often correlated, as they capture overlapping aspects of the same underlying phenomenon. Principal Component Analysis (PCA) is a dimensionality reduction technique that mitigates the effect of double-counting such correlated indicators in an index. PCA transforms the original indicators $X_1, X_2, \ldots, X_n$ into a new set of uncorrelated variables called principal components (PCs), denoted by $Z_1, Z_2, \ldots, Z_n$, where each component $Z_k$ accounts for the $k$-th largest proportion of variance in the dataset.

PCA produces two key outputs: the \textit{loadings} $a_{ki}$, which represent the contribution of the original variable $X_i$ to the principal component $Z_k$, and the \textit{eigenvalues} $\lambda_k$, which quantify the amount of variance explained by each component $Z_k$. Each principal component $Z_k$ can be expressed as a linear combination of the original indicators:

\[
Z_k = \sum_{i=1}^{n} a_{ki} X_i
\]

Typically, the first few principal components capture most of the variance in the data. These components are therefore especially useful for index construction, as they summarize essential information in fewer dimensions while avoiding redundancy.
Principal components can be directly used as indices. For example, \cite{Keyantash2024} defined the Aggregate Drought Index (ADI) as the normalized first principal component of six indicators: precipitation, evapotranspiration, streamflow, reservoir storage, soil moisture content, and snow water content. This approach has been adopted in several studies \cite{Topcu2022, Norouzi2012, Arabzadeh2016}. The weight of indicator $X_i$ is calculated as \cite{Salvati2009}:

\begin{equation}
w_i = \frac{\sum\limits_{k=1}^{n} a_{ki} \lambda_k}{\sum\limits_{k=1}^{n} \sum\limits_{l=1}^{n} a_{kl} \lambda_k}
\label{eq:pcabased}
\end{equation}

Equation \ref{eq:pcabased} is used to construct drought vulnerability indices \cite{Kim2021}. A variation of \ref{eq:pcabased} is to use the ratio of the percentage of variation explained by a component to the minimum variation explained among all of the components \cite{Mainali2017}, ignoring loadings. 

Another use of PCA is to select drought indicators for the final index \cite{Tanaya2025}. Typically, indicators with low loading values are considered unimportant and excluded from the index. However, excluding indicators is not a concern if PCs are used directly as the index, since unimportant indicators have low loadings, thereby reducing their contributions to the PCs.

\subsection{Data Envelopment Analysis}

Data Envelopment Analysis (DEA) is a non-parametric method to benchmark multiple systems. Originally, DEA was developed to evaluate the relative efficiency of decision-making units (DMUs) that use multiple inputs to produce multiple outputs \cite{Charnes1978}. Suppose there are $m$ comparable systems (DMUs), each labeled by $k=1, \ldots, m$. Each DMU $k$ uses $q$ inputs $z_{1k}, \ldots, z_{qk}$ to produce $s$ outputs $y_{1k}, \ldots, y_{sk}$. The efficiency score of DMU $k$ is defined as the ratio of the total weighted outputs to the total weighted inputs:

\[
\theta_k = \frac{\sum_{r=1}^{s} u_r y_{rk}}{\sum_{i=1}^{q} v_i z_{ik}}
\]

\noindent where $u_r$ and $v_i$ are non-negative weights assigned to the outputs and inputs, respectively. Since the input and output data are fixed, $\theta_k$ depends only on the selected weights. DEA determines the optimal weights that maximize $\theta_k$ subject to the constraint that the efficiency score for all DMUs does not exceed 1, i.e.,

\[
\frac{\sum_{r=1}^{s} u_r y_{rj}}{\sum_{i=1}^{q} v_i z_{ij}} \leq 1 \quad \text{for all } j = 1, \ldots, m, \quad u_r, v_i \geq \varepsilon > 0
\]

Let $\theta_k^*$ denote the optimal efficiency score for DMU $k$. Then, DEA suggests that DMU $k$ is more efficient than DMU $\ell$ if $\theta_k^* > \theta_\ell^*$. This method allows us to compare multiple systems by solving the following linear programming problem for each DMU $k$:

\[
P_k = \max \left\{ \theta_k \,\middle|\, \frac{\sum_{r=1}^{s} u_r y_{rj}}{\sum_{i=1}^{q} v_i z_{ij}} \le 1 \; \forall j, \; u_r, v_i \ge \varepsilon \right\}
\]

The DEA approach is non-parametric because it constructs the efficient frontier directly from the data without assuming a specific functional form. It is also unit-invariant, meaning that the results do not depend on the scale or units of the input and output variables. We can utilize the DEA to benchmark the risk or resilience of multiple geographical regions and communities.

In Algorithm \ref{alg:dea}, an output-oriented DEA approach is proposed to derive indicator weights for constructing the index. DMUs represent systems, such as geographical regions or communities, where each output is an indicator, and the total of inputs is one. All indicators represent desirable attributes such as resilience or adaptive capacity. DEA evaluates the relative performance of each DMU by assigning the most favorable weights to its indicators, subject to the condition that no DMU can have an efficiency score exceeding one under those same weights. The linear problem of DMU $k$ is defined as:

\[
P_k = \max \left\{ {\sum_{i=1}^{n} w_{ik} x_{ik}} \,\middle|\, {\sum_{i=1}^{n} w_{ik} x_{ij}} \le 1 \; \forall j, \; w_{ik} \ge \varepsilon \right\}
\]

Since the input data consists only of outputs (i.e., benefit-type indicators), this model uses a dummy input of one and treats the indicators as outputs. By solving this linear program for each DMU and averaging the resulting weights across all units, the algorithm derives a global set of indicator weights.

\begin{algorithm}
\caption{DEA-Based Weight Calculation}
\begin{algorithmic}[1]

\FOR{$k = 1$ to $m$}
    \STATE Define linear program $P_k$ for system $k$
    \STATE Define variables $w_{ik} \geq 0$ for $i = 1, \ldots, n$
    \STATE Solve problem $P_k$ to optimality and find optimal weights $w_{ik}^*$  
    \ENDFOR
\STATE Compute average weight $w_i = \frac{1}{m} \sum_{k=1}^{m} w_{ik}^*$ for all $i$
\STATE Normalize $w_i = \frac{w_i}{\sum_{j=1}^{n} w_j}$ for all $i$
\STATE Return $\{w_1, \ldots, w_n\}$
\end{algorithmic}
\label{alg:dea}
\end{algorithm}

\subsection{Criteria Importance through Intercriteria Correlation (CRITIC)}
The CRITIC method is a multi-criteria decision-making tool used to objectively determine the weights of different criteria when evaluating alternatives \cite{Diakoulaki1995}. The CRITIC method is particularly useful when the goal is to avoid redundancy in index construction while preserving the discriminatory power of diverse indicators. Algorithm \ref{algo:critic} summarizes the steps to calculate index weights. In the process, the conflict score measures dissimilarity among indicators using Pearson correlation coefficients $r_{ij}$ between each pair of $X_i$ and $X_j$.   

\begin{algorithm}
\caption{CRITIC-Based Indicator Weighting}
\begin{algorithmic}[1]
\STATE Normalize the data if needed
\STATE Compute standard deviations $ \sigma_i \quad \text{for } i = 1, \ldots, n $
\STATE Compute Pearson correlation matrix: 
$R=\text{corr\_coef}(X)$
\STATE Compute conflict for each indicator:$C_i=\sum_{j=1}^{n} \left(1 - |r_{ij}|\right)$
\STATE Compute information measure $I_i=\sigma_i \cdot C_i$
\STATE Normalize weights: $w_i=\frac{I_i}{\sum_{j=1}^{n} I_j}$
\STATE Return $\{w_1, \ldots, w_n\}$
\end{algorithmic}
\label{algo:critic}
\end{algorithm}

\section{Computational Experiments}

To evaluate the performance of the four methods given above, a series of Monte Carlo simulations was conducted. The primary objective of the simulations was to identify scenarios in which each method performed best. Each simulation involved creating a dataset with 20 systems ($m=20$) and 5 indicators ($n=5$), followed by normalization of all indicator values to the range $[0, 1]$ using min-max scaling. Several scenarios were designed to reflect a range of statistical properties relevant to index development.

In the Normal Scenario, indicator values were generated from independent normal distributions with $\mu=1$ and $\sigma=1$. In the Normal Mixed Scenario, the first three indicators were assigned higher variance ($\sigma=2$) to simulate differences in variability. This case also highlights a concern regarding the min-max scaling. The Normal Correlated scenario introduced a correlation (0.99) among the first three indicators. This scenario was intended to examine the behavior of the methods in the presence of redundant indicators. 

The Systemic Correlated scenario introduced a structure in which the means of the indicators were randomly determined from a triangular distribution such that $\mu(X_1)< \ldots<\mu(X_5)$. Random variables were sampled from multivariate standard variables to obtain correlated values between the first three indicators. The samples were then transformed into triangularly distributed indicators using inverse cumulative distribution sampling. This setting aimed to minimize the effect of min-max scaling on the variances and demonstrate the impact of the indicator distribution on the weights.

Each simulated dataset was used to compute composite indices using the weighting methods: variance-based (VAR), entropy-based (ENT), principal component analysis (PCA), CRITIC, and data envelopment analysis (DEA). For each method, indicator weights were derived and applied to compute a weighted average index for each system for 100 random iterations. For the correlated scenarios, PCA was run with one component in correlated scenarios and with three components in others. 

\section{Results}

\begin{figure}[htbp]
\centerline{\includegraphics[width=0.48\textwidth]{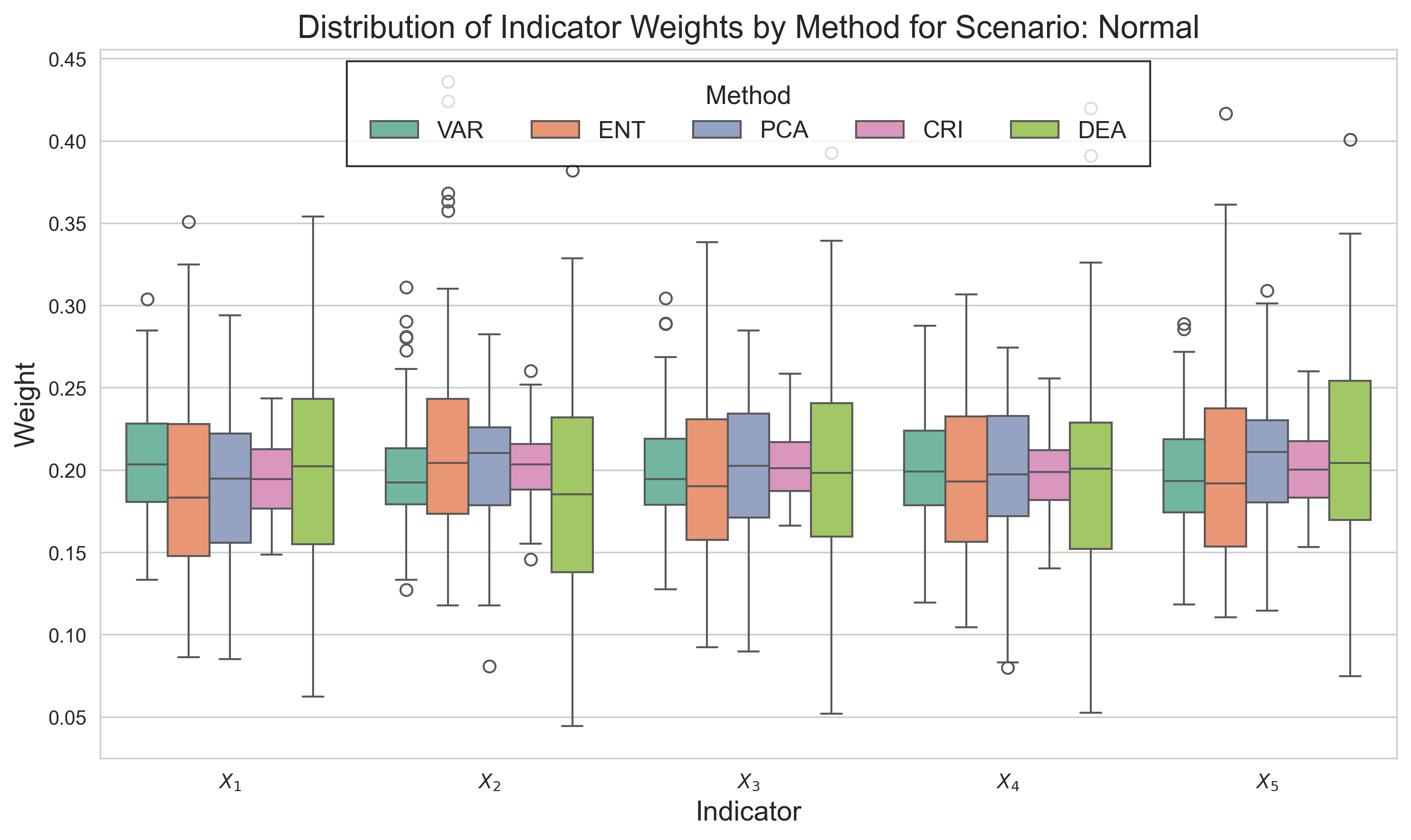}}
\caption{Boxplot of indicator weights under the Normal Scenario.}
\label{fig:normal}
\end{figure}

Fig. \ref{fig:normal} presents the boxplots of indicator weights assigned by the methods under the Normal scenario. All methods yielded a weight of 0.2 for all indicators as anticipated. DEA exhibited the greatest variability in assigned weights due to optimization, followed by ENT, whereas CRITIC consistently assigned lower, more stable weights. PCA and VAR yielded balanced and symmetric distributions across indicators.

\begin{figure}[htbp]
\centerline{\includegraphics[width=0.48\textwidth]{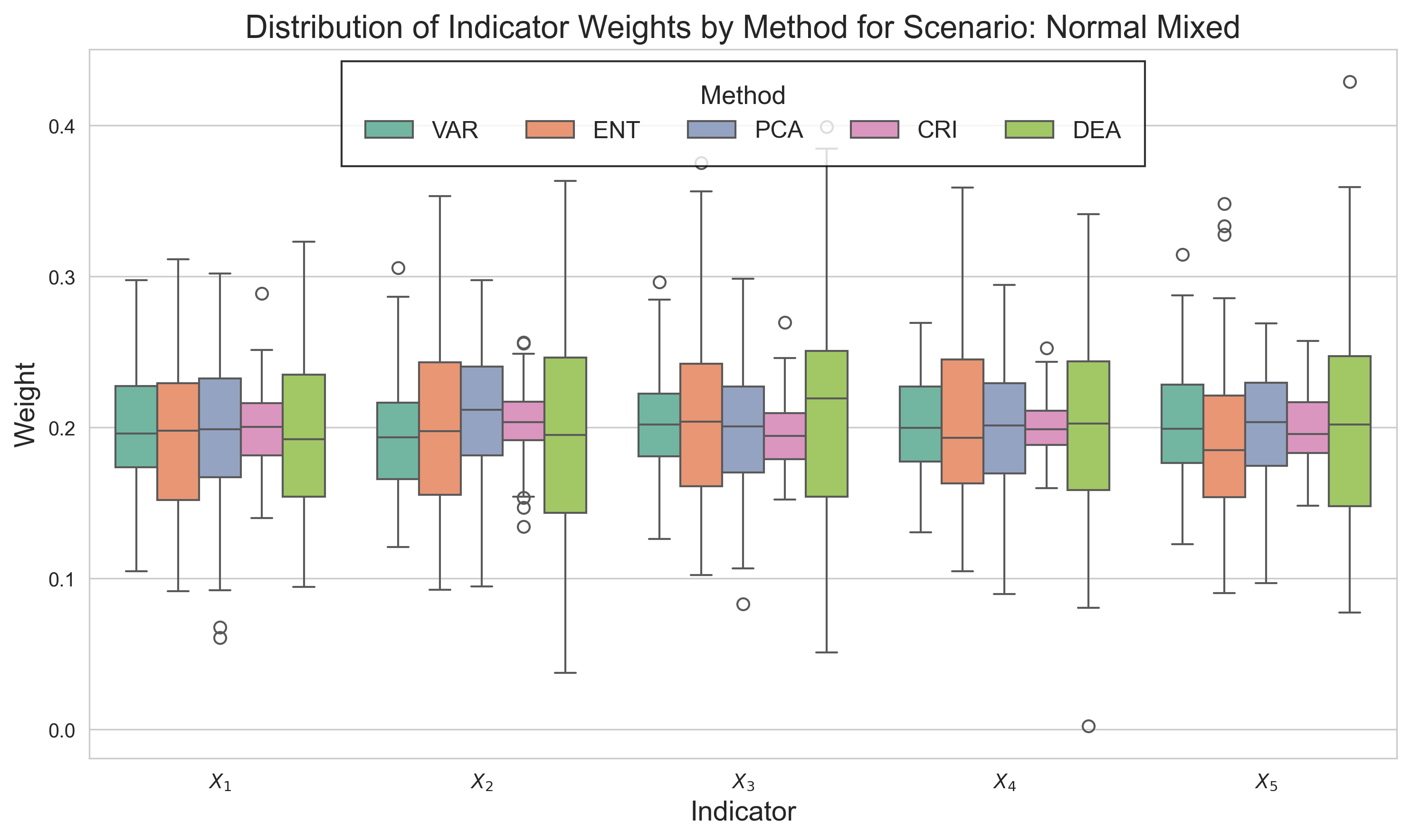}}
\caption{Boxplot of indicator weights under the Normal-Mixed simulation scenario.}
\label{fig:normal_mixed}
\end{figure}

Under the Normal-Mixed simulation scenario in Fig.~\ref{fig:normal_mixed}, all methods assigned relatively balanced weights across the five indicators, similar to the results observed in the Normal scenario. This is somewhat unexpected, as both VAR and ENT were anticipated to assign differentiated weights in response to varying levels of indicator variability, which was not observed in the results. A likely explanation is the use of min-max normalization applied to all indicators prior to weight computation. This scaling compresses each indicator’s values into the same range $[0,1]$, effectively reducing differences in raw variance and limiting the expression of information diversity (entropy) across indicators. As a result, the normalized indicators exhibit similar statistical properties, preventing VAR from emphasizing high-variance dimensions and ENT from detecting significant differences in distributional uncertainty. As a result, both methods yielded weights that are approximately uniform across the indicators despite underlying differences in the raw data. Given that min-max scaling is frequently used in index development, researchers should be mindful of this consideration. Again, CRITIC assigned lower, more concentrated weights, and DEA produced the most dispersed weight distributions.

\begin{figure}[htbp]
\centerline{\includegraphics[width=0.48\textwidth]{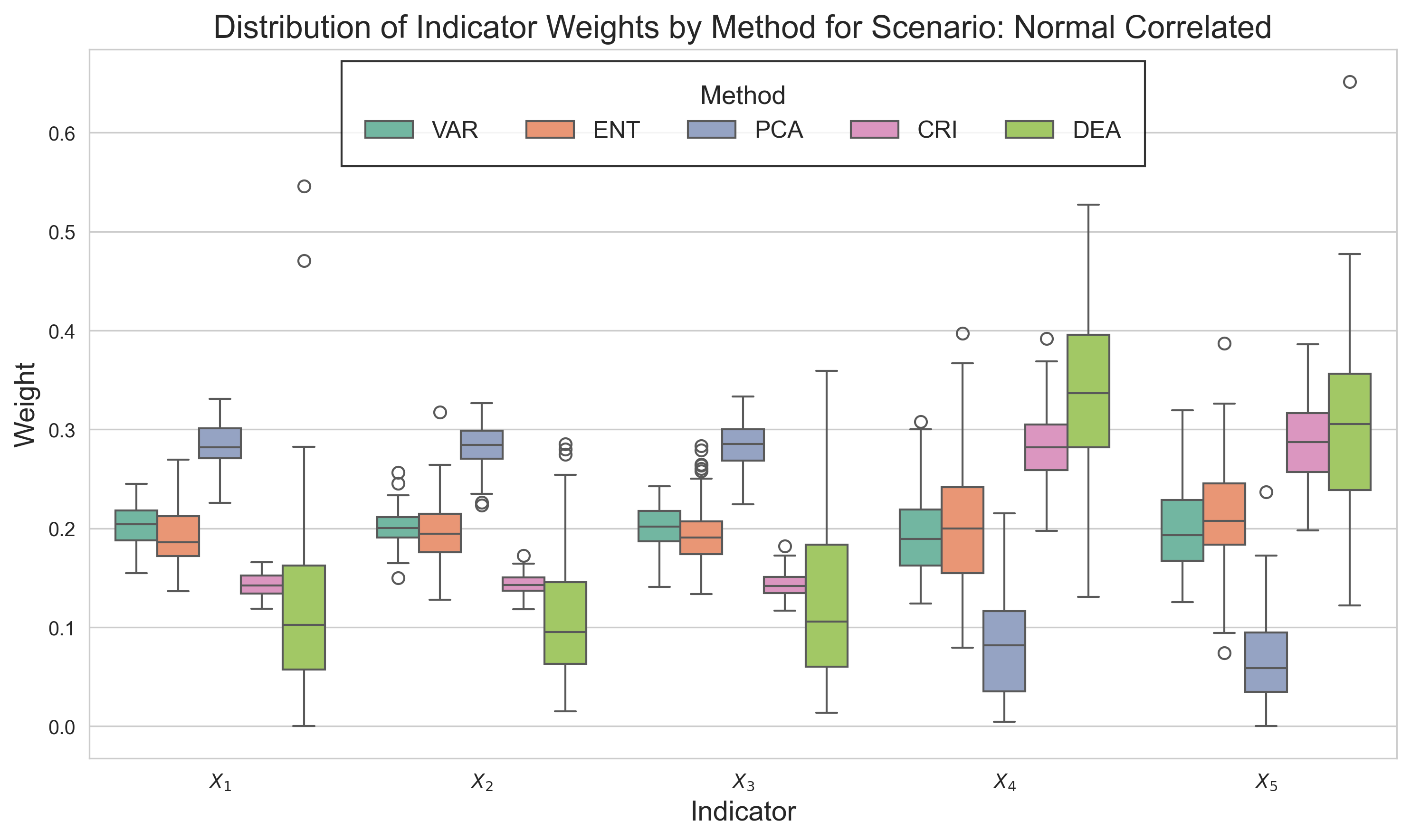}}
\caption{Boxplot of indicator weights under the Normal-Correlated simulation scenario.}
\label{fig:normal_cor}
\end{figure}

In the Normal-Correlated scenario (Fig.~\ref{fig:normal_cor}), CRITIC and DEA exhibited distinct behaviors in how they allocated weights across indicators, particularly in response to the high correlation among the first three indicators ($X_1$, $X_2$, and $X_3$). CRITIC assigned consistently lower weights to $X_1$, $X_2$, and $X_3$, with narrower distributions compared to those of $X_4$ and $X_5$. Among all methods, CRITIC produced the lowest variation for weight assignments. This is expected, as CRITIC relies on both the standard deviation (to capture contrast) and a conflict measure that penalizes indicators based on their redundancy with others. Because $X_1$, $X_2$, and $X_3$ were intentionally designed to be highly correlated, their conflict values—computed as the sum of $1 - |r_{ij}|$ with all other indicators—were low. As a result, even if they exhibited high variance, their informational contribution was considered redundant, leading CRITIC to assign them reduced weights. In contrast, the less-correlated indicators $X_4$ and $X_5$ were rewarded for their greater informational distinctiveness.

DEA uses a fundamentally different approach than CRITIC. Rather than applying a global weighting scheme, DEA derives weights endogenously for each DMU to maximize its relative efficiency. Since $X_1$, $X_2$, and $X_3$ are correlated and thus contribute little to distinguishing DMUs, DEA often emphasizes less correlated indicators, which provide greater discriminatory power when comparing DMUs. This behavior resulted in high weights assigned to $X_4$ and $X_5$ across simulations for the Normal Correlated-Variance scenario. Overall, DEA also had the most extensive range of weight distributions among all methods.

For this scenario, PCA was run for a single factor. PCA behaved in the opposite direction to DEA and CRITIC. Correlated indicators essentially measure similar underlying phenomena (in our case, $X_1$, $X_2$, and $X_3$). Therefore, PCA identifies this shared variance as the dominant pattern in the data, which becomes the first principal component. Since these three indicators contribute almost equally to this dominant pattern, they receive similar, high weights compared to other indicators when \ref{eq:pcabased} is used to calculate the weights. This demonstrates both a strength and a potential limitation of PCA for index weighting. PCA is helpful for identifying the dominant variance structure, but it might underweight indicators of conceptually critical, yet statistically independent, dimensions. 

\begin{figure}[htbp]
\centerline{\includegraphics[width=0.48\textwidth]{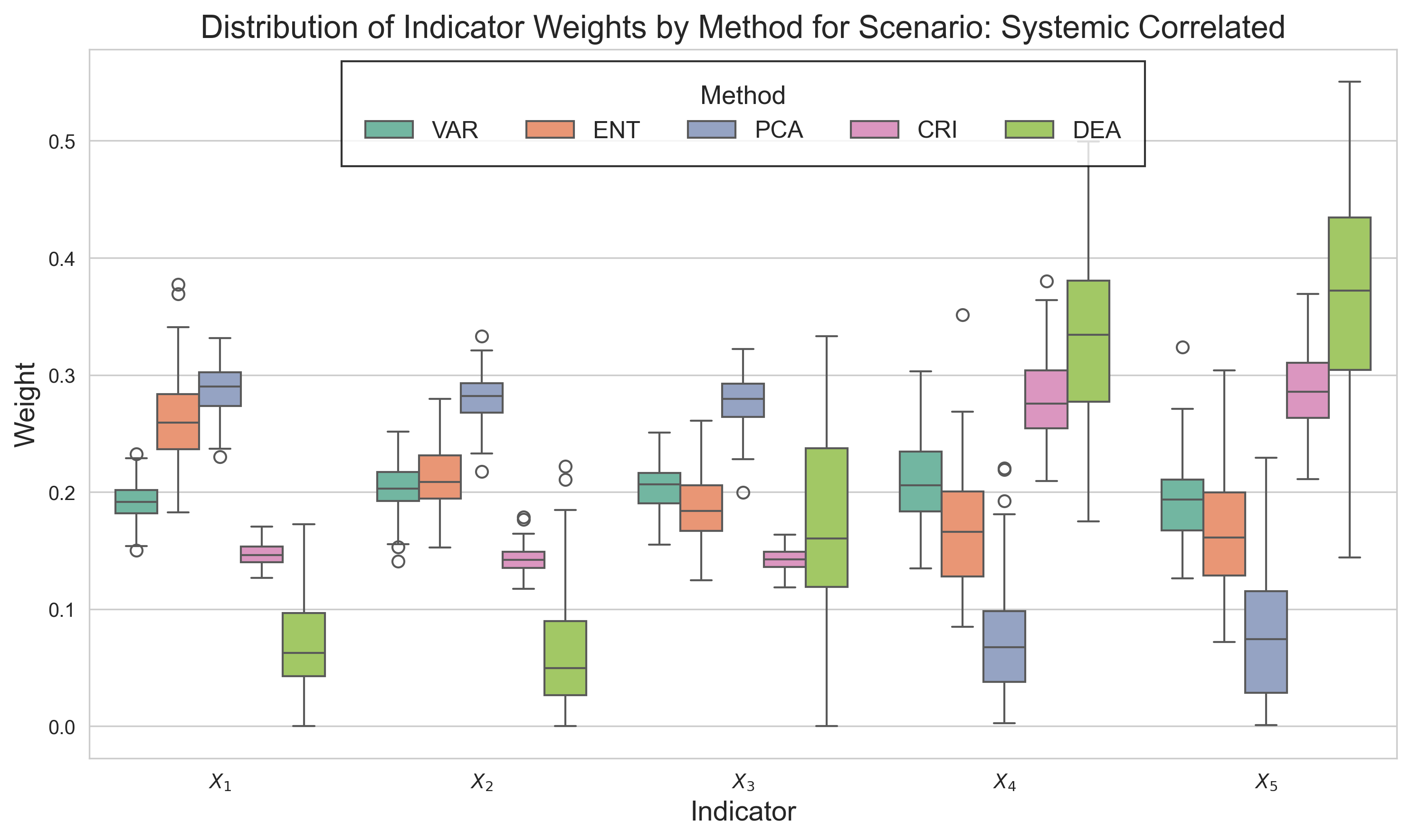}}
\caption{Boxplot of indicator weights under the Systemic Correlated simulation scenario.}
\label{fig:sys_cor}
\end{figure}

Fig.~\ref{fig:sys_cor} presents the results of the Systemic Correlated scenario. The results show a clear divergence among the methods. In this case, because the indicator distributions were not symmetric, the indicator means appeared to influence the assigned weights, even after min-max scaling. This effect was most pronounced in DEA, which assigned the largest weights to $X_4$ and $X_5$, the independent indicators with the largest means. Interestingly, DEA also assigned a relatively high weight to $X_3$, despite its strong correlation with $X_1$ and $X_2$. These outcomes are consistent with the behavior of DEA described earlier: it flexibly assigns weights to maximize DMU-level efficiency, often favoring indicators that offer distinguishing information.

Unlike in previous scenarios, ENT produced more differentiated weights across indicators. This result reflects the fact that the indicators had distinct distributional shapes; greater uncertainty results in higher entropy, and therefore higher weights. In contrast, PCA and CRITIC remained indifferent to indicator means and exhibited similar behavior to that seen in the Normal Correlated scenario. VAR also showed a similar behavior, assigning relatively uniform weights across indicators.

\section{Conclusions}
The simulation results across multiple scenarios demonstrate that the choice of indicator aggregation method has a significant impact on index construction outcomes, particularly under conditions of correlation, distributional asymmetry, and scale transformation. While the variance-based, entropy-based, and principal component analysis methods are computationally efficient and widely used, they can be highly sensitive to changes in the distributional properties of indicators, especially after normalization and scaling. Researchers using these methods must carefully assess whether scaling and indicator transformations alter variance or information content in a way that influences the resulting weights.

CRITIC offers a valuable alternative, as it explicitly accounts for both standard deviation and inter-indicator correlation. This allows CRITIC to handle correlated indicators more effectively and produce more stable weights in scenarios with overlapping information content. However, CRITIC remains sensitive to the statistical structure of the data and requires careful interpretation when applied to highly skewed or nonlinear distributions.

DEA exhibited the greatest flexibility by assigning unit-specific weights that maximize relative performance. This adaptive behavior makes DEA particularly powerful for identifying comparative advantages in heterogeneous systems. Nevertheless, DEA is computationally demanding, especially in large datasets, and it requires significant expertise to implement.

Overall, the findings highlight that researchers must explicitly justify their choice of weighting method, considering how assumptions about distribution, correlation, and scaling may impact both the interpretability and robustness of composite indices.

\section*{Acknowledgment}

This material is based upon work supported by the National Science Foundation under Grant No. 2039506 and Belmont Forum Project DR32019-Management of Disaster Risk and Societal Resilience. Any opinions, findings, and conclusions or recommendations expressed in this material are those of the author(s) and do not necessarily reflect the views of the National Science Foundation.

\bibliographystyle{IEEEtran}
\bibliography{drought}

\end{document}